\documentclass[twocolumn,prl,showpacs, amsmath, amssymb, superscriptaddress,aps,groupedaddress]{revtex4-1}
\usepackage{graphicx}
\usepackage{epstopdf}
\usepackage{float}
\usepackage{amsmath}
\begin{document}
\title{Modal makeup of transmission eigenchannels}
\author{Zhou Shi$^1$ and Azriel Z. Genack$^{1,2}$}
\affiliation{$^1$Department of Physics, Queens College of The City University of New York, Flushing, NY 11367, USA
\\$^2$Graduate Center of The City University of New York, New York, NY 10016, USA}
\date{\today}

\begin{abstract}
Transmission eigenchannels and quasi-normal modes are powerful bases for describing wave transport and controlling transmission and energy storage in disordered media. Here we elucidate the connection between these approaches by expressing the transmission matrix (TM) at a particular frequency as a sum of TMs for individual modes drawn from a broad spectral range. The wide range of transmission eigenvalues and correlation frequencies of eigenchannels of transmission is explained by the increasingly off-resonant excitation of modes contributing to eigenchannels with decreasing transmission and by the phasing between these contributions. 
\end{abstract}
\maketitle

A random speckle pattern of intensity forms within a disordered medium illuminated by a monochromatic wave as a result of the interference of innumerable scattered partial waves. This might appear to expunge any trace of the incident wave. However, there is a wide range of transmission coefficients for different incident wavefronts \cite{1,2,6,3,4}. This reflects the correlation in the transmitted flux which engenders enhanced fluctuations of transmission and conductance while suppressing average transmission of multiply scattered waves \cite{2a,1a}. Correlation in the TM also determines the degree to which the transmitted wave can be controlled by manipulating the incident wave \cite{13}. Recent measurements of the TM have shown that the incident wavefront can be adjusted to focus \cite{7,11,12,9a,28,29} and image \cite{14,15} waves through opaque samples as well as to enhance or suppress transmission \cite{8,9,10,16}.

The elements of the TM {\it t} in multichannel waveguides are the transmission coefficients $t_{ba}$ of the field between the {\it N} incident and outgoing channels, {\it a} and {\it b}, respectively. The channels may be the {\it N} orthogonal modes of the empty waveguide leading to and from the random sample. Wave propagation in such structures is analogous to electronic conduction in resistors between perfectly conducting leads at zero temperature at which inelastic electron scattering is frozen out. 

The TM was originally introduced to describe the scaling and fluctuations of electronic conductance in disordered systems in which the wavefunction is temporally coherent\cite{1,2a}. The conductance in units of the quantum of conductance $(e^2/h)$, {\textsl g}, is equivalent to the classical transmittance, $\it T$, and equals the sum of the eigenvalues $\tau_n$ of the matrix product $tt^\dagger$, ${\textsl g}=\it T=\sum_{n=1}^N \tau_n$ \cite{2b}. The Anderson transition from conducting to insulating samples occurs when the average value of ${\textsl g}$ in a random ensemble of sample realizations falls below unity \cite{19}. 

Transport in open random systems may also be described in terms of solutions of the wave equation. These are quasi-normal modes or resonances for classical waves or eigenstates in quantum systems \cite{19a}. Such modes exist for vibrational, electromagnetic or electronic degrees of freedom. The Anderson localization transition reflects the changing spatial extent of modes as a parameter of the medium or the incident wave changes. This is charted in the Thouless number $\delta$, which is essentially the ratio between the average width and spectral spacing of modes \cite{17a}. For diffusive waves, $\delta=\textsl{g}$. This reflects the close connection between modes and transmission eigenchannels. The interplay between channels and modes might explain the broad range of transmission eigenvalues and their spectral correlation and might suggest an approach to efficiently exciting selected modes within the interior of a sample. 

In this Letter, we show how transmission eigenchannels, which are defined in terms of waves at the incident and output planes of the sample, are constructed from modes, which represent the excitation within the medium. Using microwave measurements and computer simulations of the TM, we show that high-transmission eigenchannels are composed of modes with central frequencies close to the frequency of the incident wave, while low transmission eigenchannels are made up of many spectrally remote modes which interfere destructively. This leads to a wide range of transmission eigenvalues and a broadening of the spectral correlation function of eigenchannels with decreasing transmission. These results show that modes can be most efficiently selected for applications to random lasing and spectroscopic sensitivity by illuminating a sample with the modal transmission eigenchannel. 

Spectra of the TM for microwave radiation propagation through a copper waveguide filled with randomly positioned alumina spheres are measured with the use of a vector network analyzer. A detailed experimental setup can be found in the Supplementary Material \cite{17aa}. We express the TM at angular frequency $\omega$ in terms of the TMs of the individual modes of the random system, 
\begin{equation}
t(\omega) = \sum_m t^m (\omega) = \sum_m t_{ba}^m\frac{\Gamma_m/2}{\Gamma_m/2+i(\omega-\omega_m)}
\end{equation}
Here, $\omega_m$ is the central angular frequency of the $m^{th}$ mode, $\Gamma_m$ is the corresponding linewidth and $ t_{ba}^m $ is the field transmission coefficient through the $m^{th}$ mode for excitation from incident channel {\it a} and detection at output channel {\it b} at $\omega=\omega_m$. In the absence of dissipation, $\Gamma_m$ equals the leakage rate of modal energy through the sample boundaries. 
\begin{figure}[htc]
\centering
\includegraphics[width=4in]{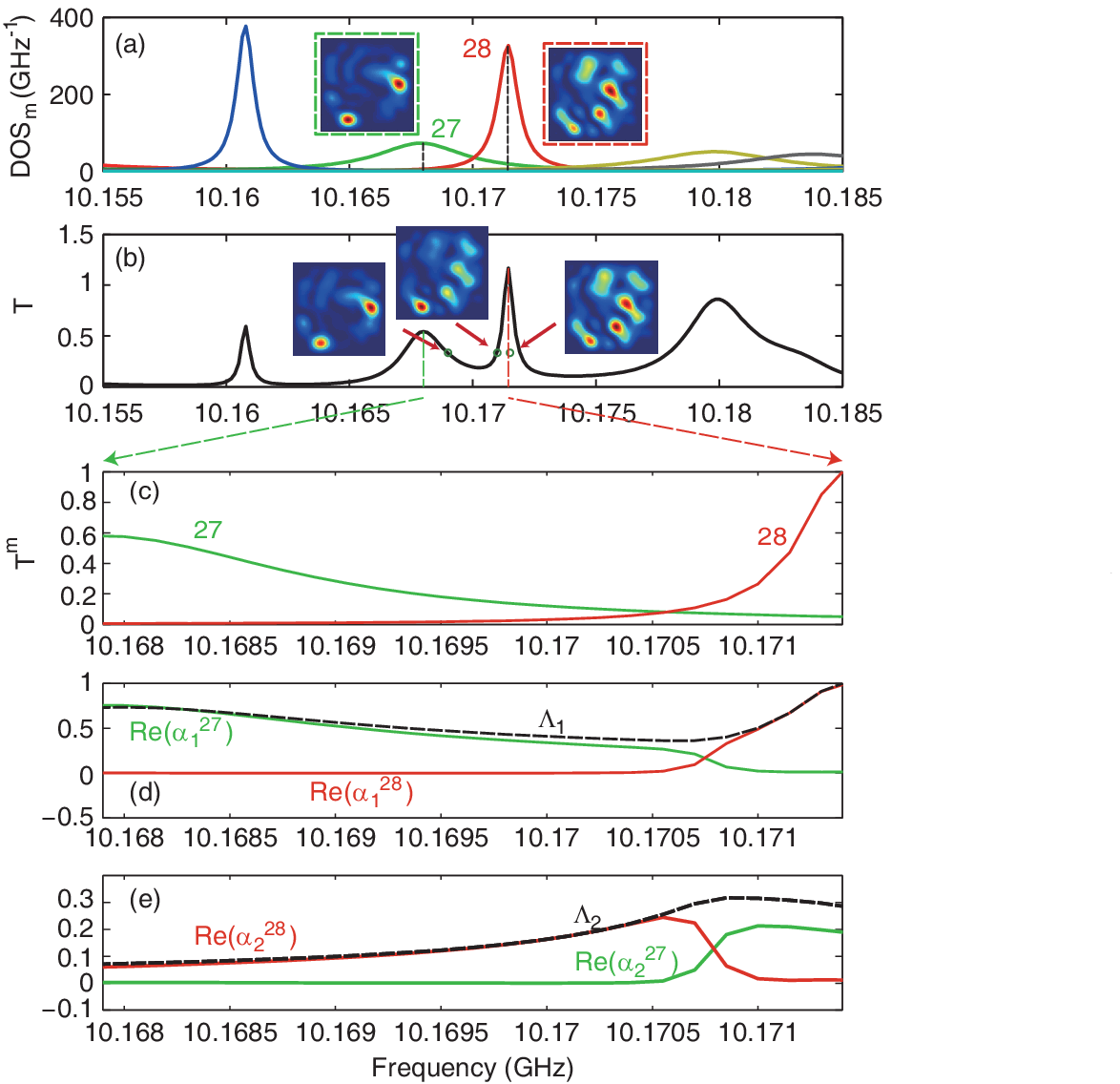}
\caption{(Color online). Modes and channels in a disordered waveguide. (a) Spectra of the density of states (DOS) of individual modes of a disorder sample drawn from an ensemble with dimensionless conductance {\textsl g} = 0.37. The speckle patterns for Modes 27 and 28 at the output surfaces are shown. The 2D sampling theorem is employed to produce the high-resolution speckle patterns. (b) Spectrum of transmittance {\it T}. The output speckle patterns for the first transmission eigenchannel at three frequencies indicated by green circles are shown. (c) Spectra of the transmittance from Modes 27 and 28 acting alone in the frequency range from the central frequency of Mode 27 to the central frequency of Mode 28. (d-e) Frequency variation of the contributions to the first and second eigenchannels from Modes 27 (green curve) and 28 (red curve).} \label{Fig1}
\end{figure} 

The central frequencies and widths of the modes are obtained both in experiments and simulations, utilizing nonlinear least-squares optimization methods to simultaneously fit multiple spectra of transmission with Eq. 1. The statistical impact of absorption upon measurements of the statistics of propagation is largely removed by subtracting a constant width equal to the absorption rate from the linewidths of each of the modes and enhancing the resonant amplitude for each modes \cite{20}. 

At a given frequency, the TM can be expressed via the singular value decomposition as $t=U\Lambda V^\dagger$ \cite{1,2}. The columns of the unitary matrices $U$ and $V$ are the singular vectors of the TM and $\Lambda$ is a diagonal matrix with singular values $\Lambda_n$ along its diagonal. The square of the elements of $\Lambda_n=U_n^\dagger tV_n$ are the transmission eigenvalues $\tau_n$. The mode transmission matrix at a given frequency can also be expressed in terms of its singular vectors, $t^m=u^m\lambda^mv^{m\dagger}$. Lower (upper) case symbols are used to represent $t^m$ ($t$). Since the modes of a system should be independent of the way they are excited, the speckle pattern at the sample output for a mode excited by a source at different points in the incident plane should differ only up to a multiplicative constant. The columns of $t^m$ would then be linearly related and $t^m$ would be of rank one. Indeed, for localized waves, for which typically $\tau_2/\tau_1\ll1$, so that the participation number of transmission eigenchannels, $M \equiv (\sum_{n=1}^N \tau_n)^2/\sum_{n=1}^N \tau_n^2$, \cite {11} is close to unity, the measured speckle pattern normalized by its peak value hardly changes with source position \cite{8}. Taking the rank of $t^m$ to be unity, we can write
\begin{equation} 
\Lambda_n = U_n^\dagger \sum_m t^m V_n = \sum_m U_n^\dagger u_1^m\lambda_1^mv_1^{m\dagger}V_n=\sum_m \alpha_n^m 
\end{equation}
In this expression, the vector inner produce $v_1^{m\dagger}V_n$ and $U_n^\dagger u_1^m$ describe the overlap of the $n^{th}$ eigenchannel and the $m^{th}$ mode at the input and output surfaces, respectively. The sum of the components of the $\alpha^m_n$ that are in phase with $U_n$ is equal to $\Lambda_n$. 

We find in both experiments and simulations that $\langle \lambda_2^m\rangle$ of the $t^m$ determined from the fitting procedure for modes, does not vanish. Typically, $\langle\lambda_2^m/\lambda_1^m\rangle\sim 0.1$ while $\langle(\lambda_2^m)^2/(\lambda_1^m)^2\rangle\sim 0.04$, indicating that the modal decomposition is not perfect. The reasons for a finite value of $\langle\lambda_2\rangle$ are not clear at present, but the effect is small enough that the qualitative nature of the relationship between modes and channels emerges from an analysis of measurements and simulations.

We present an example of the modal analysis of the measured transmission eigenchannels of a random sample in Fig. 1. The Lorentzian density of states for several individual modes, DOS$_m$, with spectral integral of unity is plotted in Fig. 1a. The modes are numbered in ascending order with increasing frequency in the measured spectrum from 10 to 10.24 GHz. The output speckle patterns for Modes 27 and 28 are shown with central frequencies indicated by the vertical black dashed line. The spectrum of {\it T} together with the output speckle patterns for the first eigenchannel, $U_1$ for the three frequencies indicated by green circles are shown in Fig. 1b. Spectra of the transmittance through individual modes, $T^m(\omega) = \sum_{a,b} |t_{ba}^m\frac{\Gamma_m/2}{\Gamma_m/2+i(\omega-\omega_m)}|^2$ for Modes 27 and 28 between the central frequencies of these two modes are shown in Fig. 1c. The contributions from these modes to the singular values $\Lambda_n$ are shown in Figs. 1d,e for {\it n} = 1,2. For localized waves, the first eigenchannel is dominated by the mode which is most nearly resonant with the incident wave. This can also be seen in the similarities between the output speckle patterns of the first transmission eigenchannel (Fig. 1b) and the output speckle patterns of the corresponding modes (Fig. 1a) \cite{21a,22}. In contrast, the second transmission eigenchannel in this configuration is seen in Fig. 1e to be made up largely of the next nearest mode.

The measured transmission eigenvalues fall exponentially with channel index $n$ for localized waves \cite{1,8}. For the sample represented in Fig. 1, the signal falls below the noise level for $n>3$. We have therefore performed numerical simulations to explore the modal composition of low transmission eigenchannels for a scale wave propagating through a 2D disordered waveguide with {\textsl g} = 0.26 \cite{17aa}.
\begin{figure}[htc]
\centering
\includegraphics[width=3.4 in]{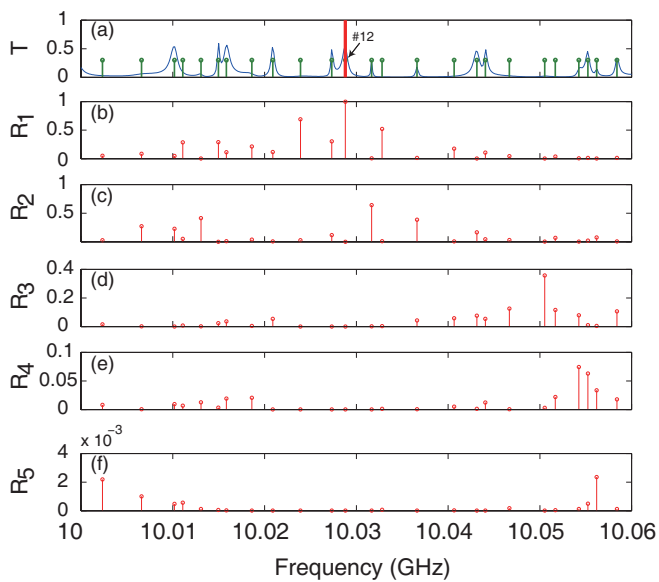}
\caption{(Color online). Simulations of the excitation of quasi-normal modes inside a random quasi-1D strip by different transmission eigenchannels. (a) Spectrum of transmittance {\it T} for a sample with {\textsl g}= 0.26. The excitation frequency in all frames is indicated by the thick line in (a), while the central frequencies of the modes are marked with thin lines. (b-f) Effectiveness of coupling into and coupling out of a mode with transmission eigenchannels with {\it n} = 1 to 5, $R^m_n$.} \label{Fig2}
\end{figure} 

The simulated spectrum of $T$ and the central frequencies of the underlying modes in a disordered sample are shown in Fig. 2a. The correlation between speckle patterns of the eigenchannels and modes due to both the input and output speckle patterns, maybe characterized by the parameter, $R^m_n=|U_n^\dagger t^mV_n|^2/|u_1^{m\dagger}t^mv_1^m|^2$. Plots of $R_n^m$ for the first 5 eigenchannels for the modes in the frequency range studied is shown in Figs. 2b-f. Modes that are close in frequency to the excitation frequency indicated by the thick red line in Fig. 2a contribute significantly to highly transmitting eigenchannels; distant modes are relatively more prominent in lower transmission eigenchannels. 

The increasing of relative contribution from spectrally remote mode to the lower transmission eigenchannels at a given frequency can explain the broadening with $n$ of the spectral correlation function of transmission eigenchannels, $C(\Delta\nu)=\langle U_n^*(\nu)U_n(\nu+\Delta\nu)\rangle$, which is shown in Fig. 3. The variation with frequency shift of both the amplitudes and phases of the off-resonant modal contributions to low transmitting eigenchannels is slow. The mode amplitudes for off-resonance modes fall inversely with frequency shift for modes far from resonance (Eq. 1). In contrast, both the amplitudes and phases of individual modes change quickly with frequency shift for nearly resonant modes which contribute appreciably to high transmission eigenchannels. 
\begin{figure}[htc]
\centering
\includegraphics[width=3.5in]{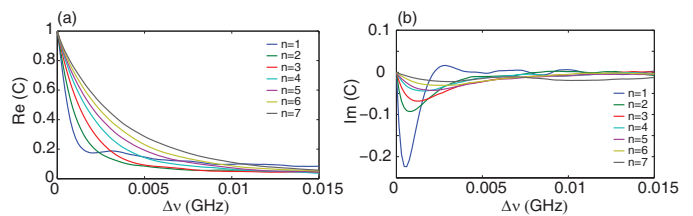}
\caption{(Color online). Frequency correlation of transmission eigenchannels in simulations. Real and imaginary parts of the correlation function are shown in (a) and (b), respectively.} \label{Fig3}
\end{figure} 

The contribution of each mode to the singular value for each transmission eigenchannel depends upon the following factors: (1) the closeness to resonance with the mode, (2) the projection of the incident wave for the eigenchannel $V_n$ upon that for the mode $v{_1}^m$, (3) the projection of the outgoing wave through a single mode $u{_1}^m$ upon the eigenchannel $U_n$, and (4) the phasing of the modal contributions. All of these factors can be combined into a single vector representing the contribution of each mode to each singular value via the vector model illustrated in Fig. 4a. The horizontal and vertical components of the vector $\alpha^m_n$ are the in- and out-of-phase components of the projection of the corresponding output speckle pattern of the $m^{th}$ mode excited by $V_n$ upon the vector $U_n$ . The vector representation of the modal contributions to the first three eigenchannels for the excitation frequency shown in Fig. 2a is shown in Figs. 4b-d. In this case, the first transmission eigenchannel is dominated by the single resonant contribution of Mode 12, whereas the second and third eigenchannels are the superpositions of several vectors associated with off-resonance modes. Each of the modal contributions is small. Moreover, the contributions are randomly phased so that transmission is further reduced by destructive interference. 

When a transmission eigenchannel is largely composed of a single mode, $\lambda_1^m$ cannot be great than unity, because transmission through a passive medium cannot be larger than the incident flux. However, when two or more modes overlap and have similar speckle patterns in transmission, they would form a single transmission eigenchannel \cite{21a}. We find then that the value of $\lambda_1^m$ can be greater than unity for one or more modes. When this occurs, the maximum value of transmission through the sample is reduced below unity by destructive interference between such modes. This is seen in the spectra of $T$ and of $\tau_1^m\equiv(\lambda_1^m)^2$ in Fig. 4e and in the vector diagram in Fig. 4f in a case in which $\lambda_1^m>1$ for two modes while $T<1$. 
\begin{figure}[htc]
\centering
\includegraphics[width=3.4in]{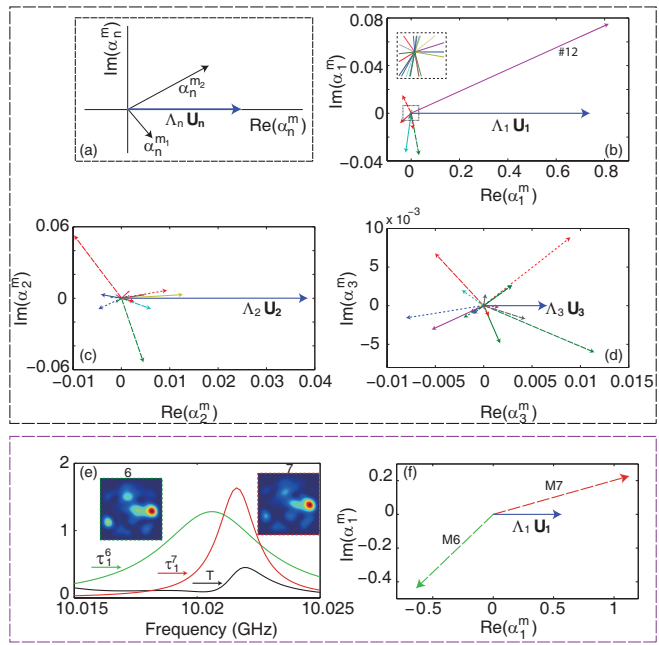}
\caption{(Color online). Vector representation of modal contribution to transmission eigenchannels. (a) Illustration of the vector model. Each vector is the contribution from the $m^{th}$ mode to the output of the $n^{th}$ transmission eigenchannel, $U_n$. The angle with the horizontal axis is the overall phase shift between the mode and eigenchannel output speckle patterns, $u_1^m$ and $U_n$, respectively. The resultant vector is equal to $\Lambda_nU_n$. (b-d) Contributions to the first, second and third eigenchannels at the excitation frequency shown in Fig. 2a from each mode in the sample found in simulations. (e) Spectrum of the transmittance {\it T} and $\tau_1^m=(\lambda_1^m)^2$ for Modes 6 and 7 for one sample realization in the experiment. (f) The vector diagram for the first transmission eigenchannel for the same sample as in (e) at the central frequency of the Mode 7.} \label{Fig3}
\end{figure} 

For diffusive waves for which typically ${\textsl g=\delta}$ modes are excited on resonance, the transmittance is dominated by approximately {\textsl g} eigenvalues \cite{1}. Since these ``open" eigenchannels are orthogonal and are a linear combination of the speckle patterns of the nearby modes, the speckle patterns of these modes are expected to differ as found in the numerical simulation \cite{21}. 

Some characteristics of the modal makeup of eigenchannels for diffusive waves might be gleaned from consideration in a case in which two modes with distinct speckle patterns make comparable contribution to transmission. We consider transmission at a frequency of 10.1707 GHz in the measured spectrum shown in Fig. 1 which lies between the central frequencies of the two modes with dissimilar speckle patterns. At this frequency, the first and second transmission eigenvalues are comparable. The corresponding vector models are shown in Fig. 5a and 5b. The vectors for the contributions of the two closest modes to the exciting frequency are seen to have positive components along the first and second eigenchannels. This suggests the possibility that for transmission of diffusive waves the vectors associated with on-resonance modes will tend to have positive projections along the resultant, $U_n$ for $n<{\textsl g}$. This would result in a large resultant and so a high value of the transmission eigenvalue \cite{1}. 
\begin{figure}[htc]
\centering
\includegraphics[width=3.4in]{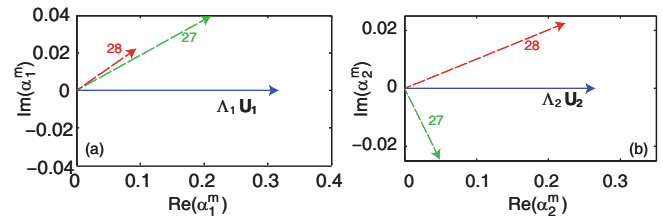}
\caption{(Color online). The vector model for the contribution to first and second eigenchannels at frequency 10.1707 GHz from Modes 27 and 28 found in measurements shown in Fig. 1.} \label{Fig5}
\end{figure} 

When two modes exhibit distinct optimal excitation patterns $v_m^1$, considerable modal selectivity can be achieved by exciting the sample on resonance with the wavefront corresponding to that of the desired mode. Exciting the sample with a sharply truncated pulse with a rise time of $\Gamma_m^{-1}$ \cite{50}, should further increase the discrimination between modes. Maximal sample excitation for various linear and nonlinear probes may be exploited to enhance the sensitivity of linear and non-linear probes and to lower the threshold of random lasers \cite{31}. Fundamental limits to the selective excitation of modes arise, however, for spectrally overlapping modes with similar speckle patterns in the circumstance that $\lambda_m^1>1$ since this would violate energy conservation.

In conclusion, we have analyzed the TM at a single frequency in terms of spectrum of modes in random media. The modal analysis of the TM explains the wide range of transmission eigenvalues in random media and the increasing correlation frequency of more weakly transmitting eigenchannels. These results suggest an approach for manipulating the incident wavefront to selectively excite specific modes with desired spectra, temporal or spatial characteristics and indicate the limits of selectivity for spectrally overlapping modes. 

We thank Arthur Goetschy and A. Douglas Stone for providing the simulation code to calculate the transmission matrix through a two dimensional disordered waveguide. The research was supported by the National Science Foundation (DMR-1207446).

\end{document}


\title{Supplementary Material \\Modal makeup of transmission eigenchannels}
\author{Zhou Shi$^1$ and Azriel Z. Genack$^{1,2}$}
\affiliation{$^1$Department of Physics, Queens College of The City University of New York, Flushing, NY 11367, USA
\\$^2$Graduate Center of The City University of New York, New York, NY 10016, USA}
\date{\today}
\maketitle

\section{Experimental Setup}

The random samples are comprised of a copper waveguide filled with randomly positioned alumina spheres of diameter 0.95 cm and refractive index 3.14 embedded in Styrofoam shells. The diameter of the copper tube is 7.3 cm and the length {\it L} = 40 cm. The volume filling fraction of the alumina spheres is 0.068 and the dimensionless conductance {\textsl g} of the sample is equal to 0.37. Polarized microwave radiation at incident point {\it a} is provided by a wire antenna connected to a vector network analyzer, which can translate freely on the input surface of the tube. The transmission coefficient at a point {\it b} is measured by another wire antenna which can also move on a grid on the output surface. The spacing of the grid is 9 mm. Spectra of the TM over a frequency range from 10 to 10.24 GHz in 801 steps are obtained from spectra of transmission coefficients $t_{ba}$ between 49 points on the input and out surfaces. The sample tube is rotated and vibrated momentarily after each measurement of the TM spectrum to produce a statistically equivalent sample realization. Though there is still considerable overlap between modes in this weakly localized system, it is still possible to analyze the TM at a given frequency into a sum of TMs for individual modes in the system. 

In our microwave measurements, only a fraction of the transmitted energy is measured since the TM is determined by measuring the transmission coefficients on a grid with a finite number of points and for a single polarization. Therefore, the measured TM is not complete. The impact of the incomplete measurement of the TM upon the probability density of eigenvalues of the TM has been recently investigated by Goetschy and Stone \cite{33}. They explored the change in the density of transmission eigenvalues from the bimodal distribution to a distribution characteristics of Gaussian random matrices as the ratio of measured channels $N^\prime$ and total number of channels $N$ on the input side $m_1=N_1^\prime/N$ and output side $m_2=N_2^\prime/N$  decreases. However, the statistics of transmission are well represented by the measured TM provided that the number of channels measured is much greater than {\textsl g} \cite{33,32}. 

\section{Recursive Green's Function Simulation}
To explore the modal composition of low transmission eigenchannels, we have performed Green's function simulations for a scalar wave propagating through locally 2D random waveguides with ideal leads attached in the longitudinal directions and perfect reflecting transverse sides. The disorder region in the waveguide is modeled by a position-dependent dielectric constant, $\epsilon(x,y) = 1+ \delta\epsilon(x,y)$ with $\delta\epsilon$ drawn randomly from the uniform distribution [-0.9,0.9]. The dielectric constant of the ideal leads matches the average dielectric constant of the disordered sample, so that the impact of both internal and external reflections at the sample boundaries is minimized. The wave equation $\triangledown^2E(x,y)+k_0^2\epsilon(x,y)E(x,y)=0$ is discretized using a 2D tight-binding model on a square grid and solved with the use of the recursive Green's function method \cite{34a}. The sample length {\it L} and width {\it W} are 550 and 50 in units of the grid spacing, respectively. Since the length of the sample greatly exceeds its width, the sample can be considered as quasi-one-dimensional. Simulations are made for 200 configurations over a narrow frequency range from 10 to 10.06 GHz in 401 steps. The product of the wave number $k_0$ in the leads and the grid spacing is set to be unity at 10 GHz. The number of propagating modes of the empty waveguide over the frequency range is $N=16$. The transmission coefficient between incoming waveguide mode {\it m} and outgoing mode {\it n} in the leads, $t_{nm}$, is obtained by projecting the Green's function between arrays of points on the input and output surfaces onto the wavefunction of the waveguide modes {\it m} and {\it n}, respectively. The sum of $N^2$ pairs of the square of the amplitude of $t_{nm}$ gives the transmittance {\it T} of the sample, of which {\textsl g} = 0.26.